\documentclass[reprint,aps,prl,showpacs,amsmath,superscriptaddress,floatfix]{revtex4-1}
\usepackage{amsmath,amsfonts,amssymb}
\usepackage{graphicx}
\usepackage[utf8]{inputenc}

\usepackage[usenames,dvipsnames]{color}
\usepackage[normalem]{ulem}

\begin{document}
\title{Sliding drops -- ensemble statistics from single drop bifurcations}
\author{Markus Wilczek}
\email[]{markuswilczek@uni-muenster.de}
\affiliation{Institute for Theoretical Physics, University of M\"unster, Wilhelm-Klemm-Str.\ 9, D-48149 M\"unster, Germany}
\affiliation{Center for Nonlinear Science (CeNoS), University of M\"unster, Corrensstr.\ 2, D-48149 M\"unster, Germany}
\author{Walter Tewes}
\affiliation{Institute for Theoretical Physics, University of M\"unster, Wilhelm-Klemm-Str.\ 9, D-48149 M\"unster, Germany}
\author{Sebastian Engelnkemper}
\affiliation{Institute for Theoretical Physics, University of M\"unster, Wilhelm-Klemm-Str.\ 9, D-48149 M\"unster, Germany}
\author{Svetlana V. Gurevich}
\author{Uwe Thiele}
\email[]{u.thiele@uni-muenster.de}
\affiliation{Institute for Theoretical Physics, University of M\"unster, Wilhelm-Klemm-Str.\ 9, D-48149 M\"unster, Germany}
\affiliation{Center for Nonlinear Science (CeNoS), University of M\"unster, Corrensstr.\ 2, D-48149 M\"unster, Germany}
\affiliation{Center for Multiscale Theory and Computation (CMTC), University of M\"unster, Corrensstr.\ 40, D-48149 M\"unster, Germany}

\date{\today}

\begin{abstract}
Ensembles of interacting drops that slide down an inclined plate show a dramatically different coarsening behavior as compared to drops on a horizontal plate: As drops of different size slide at different velocities, frequent collisions result in fast coalescence. However, above a certain size individual sliding drops are unstable and break up into smaller drops. Therefore, the long-time dynamics of a large drop ensemble is governed by a balance of merging and splitting. We employ a long-wave film height evolution equation and determine the dynamics of the drop size distribution towards a stationary state from direct numerical simulations on large domains. The main features of the distribution are then related to the bifurcation diagram of individual drops obtained by numerical path continuation. The gained knowledge allows us to develop a Smoluchowski-type statistical model for the ensemble dynamics that well compares to full direct simulations. 
\end{abstract}

\maketitle

\paragraph{Introduction}
The coarsening of small-scale structures as, for instance, clusters, crystals, drops or quantum dots, into larger ones is a fundamental and widely investigated physical process common in nature and technology \cite{Bray1994ap,Nepo2015crp}. Amongst the first investigations of coarsening dynamics is Ostwald's work on the growth of larger crystals or particles in solution at the expense of smaller ones \cite{Ostwald1896Lehrbuch}. Often, these processes of \textit{Ostwald ripening} can be described by scaling laws for the time evolution of typical length scales. The power law scaling for cluster growth was explained by Lifshitz and Slyozov \cite{lifshitz1961kinetics} and independently by Wagner \cite{wagner1961theorie}. In their derivation, the dynamics of the individual objects is related to the dynamics of the entire ensemble.

A particular soft matter example is an ensemble of liquid drops on a solid substrate which naturally exhibits coarsening. The statistical description of condensing and coarsening drops on horizontal substrates, i.e., their evolution towards equilibrium, was addressed by Meakin and coworkers in terms of particle based statistical models \cite{meakin1992droplet} and also by Smoluchowski-type integro-differential equations for volume distribution functions \cite{Smoluchowski1916}. They also consider the case of inclined substrates, where the drops are initially pinned and then depin at a critical volume in an avalanche process. 
The coarsening and migration of liquid drops on horizontal substrates was also addressed in detail for the one- (1D) \cite{GrWi2009pd,KiWa2010jem,Kita2014ejam,glasner2003coarsening} and two-dimensional (2D) case \cite{PiPo2004pf,otto2006coarsening} employing a lubrication or long-wave model \cite{Oron1997Long,Thie2007}. The relation to Ostwald ripening is also discussed in \cite{GORS2009ejam}.

Here, we analyze the coarsening dynamics of liquid drops that due to gravitation slide down a plate of fixed inclination. The lateral motion of the drops with respect to each other depends strongly on differences in drop size resulting in a fast relative transport that facilitates coarsening, i.e., the coalescence of smaller drops into larger ones. At the same time, large drops above a certain critical size are unstable with respect to break-up into smaller ones, due to the so-called \textsl{pearling instability}. Similarly, drops of fixed size are unstable above a critical substrate inclination \cite{Podgorski2001Corners,EWGT2016PRF}. We investigate the interplay of the accelerated coarsening and the pearling instability and elucidate the resulting statistical properties of large ensembles of sliding drops. To this end, we employ a long-wave film height evolution equation and conduct large-scale direct numerical simulations (DNS) of sliding drop ensembles to extract the dynamics of statistical measures like the drop size distribution. Next, the resulting stationary distribution of the ensemble is related to the bifurcation diagram and stability properties of individual drops obtained by numerical path continuation techniques \cite{DWCD2014ccp}. Finally, we merge the numerically obtained single-drop information including several scaling laws and develop an augmented Smoluchowski coagulation equation as simple statistical model that describes the dynamics of the drop size distribution.

\paragraph{Modelling and Numerical Implementation}
\begin{figure*}[tp]
 \centering
 \includegraphics{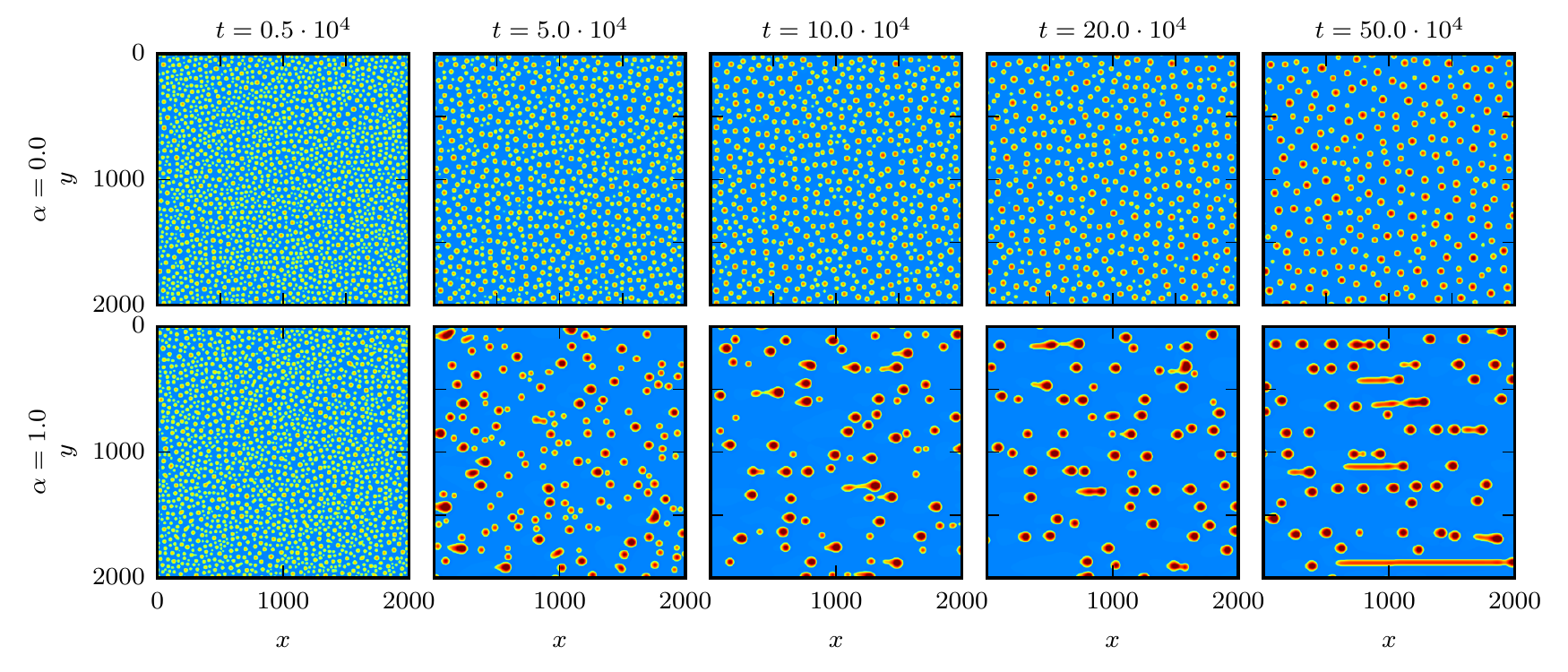}
 \caption{Snapshots at different times $t$ from DNS of the model \eqref{eq:ThinFilmEquationSliding} for (top) horizontal $\alpha=0$ and (bottom) inclined $\alpha=1$ substrates. The drops slide from left to right and only one quarter of the computational domain is shown. The supplementary material contains the corresponding videos.}
 \label{fig:fig1_manysliding_timeseries}
\end{figure*}
A non-dimensional long-wave equation is used to model the time evolution of the height profile $h(x,y,t)$ that describes drops of a simple liquid on a partially wetting substrate, cf.~\cite{EWGT2016PRF}:
\begin{equation}
\label{eq:ThinFilmEquationSliding}
 \partial_t h = - \nabla \cdot \left[ \frac{h^3}{3} \nabla \bigr(\Delta h +\Pi(h) \bigr) + \frac{h^3}{3}G\begin{pmatrix} \alpha \\ 0\end{pmatrix}\right].
\end{equation} 
The model accounts for the surface tension of the liquid via a Laplace pressure, substrate-liquid interactions such as wettability via a Derjaguin (or disjoining) pressure $\Pi(h) = - \frac{1}{h^3} + \frac{1}{h^6}$ and for the lateral driving where $G$ and $\alpha$ are a non-dimensional gravity parameter and the inclination angle of the substrate, respectively \footnote{Starting from a dimensional form $\tilde \Pi(\tilde h) = - A/\tilde h^3 + B/\tilde h^6$ for the disjoining pressure, we employ scales $h_\mathrm{eq} = (B/A)^{1/3}$ for height, $l_0 = \sqrt{3}h_\mathrm{eq}/\sqrt{5}\theta_\mathrm{eq}$ for lateral lengths, and $t_0 = 9 \eta h_\mathrm{eq}/25\gamma \theta_\mathrm{eq}^4$ for time. Then $\theta_\mathrm{eq} = \sqrt{3 A/5 \gamma h_\mathrm{eq}^2}$ is the equilibrium contact angle  and $G = 3\rho g h^2_\mathrm{eq}/5 \gamma \theta_\mathrm{eq}^2$ is the gravitation number. Here, we use $G = 10^{-3}$.}. The employed Derjaguin pressure \cite{pismen2001nonlocal} results in the presence of a thin adsorption layer in the whole domain on which the drops slide. DNS of this model are conducted on a large spatial domain $\Omega = [0,4000]\times[0,4000]$ with periodic boundary conditions using a finite-element method on a quadratic mesh with bilinear ansatz functions and a 2nd-order implicit Runge-Kutta scheme for time-stepping, implemented using the DUNE PDELab framework \cite{bastian2010generic,bastian2008genericI,bastian2008genericII} (for more numerical details see \cite{EWGT2016PRF}). 

\paragraph{Properties of the Drop Ensembles}
Figure~\ref{fig:fig1_manysliding_timeseries} presents simulation snapshots at different times $t$ that contrast coarsening on a horizontal ($\alpha = 0$, top row) and an inclined ($\alpha = 1.0$, bottom row) substrate. Up to $t\approx 0.5\cdot10^4$, the coarsening proceeds very similarly in both cases, however, at non-zero inclination the later stages are dominated by a faster coarsening process that results in larger drop sizes. This continues until a certain time  ($t_\mathrm{c}\approx 6.5\cdot 10^4$ at $\alpha = 1.0$), after which the typical drop size hardly increases further, because the pearling instability breaks up all drops above a certain volume. Only at very late stages of the simulation, a tendency to form large elongated drops can be noted.

\begin{figure}[htbp]
\center
 \includegraphics{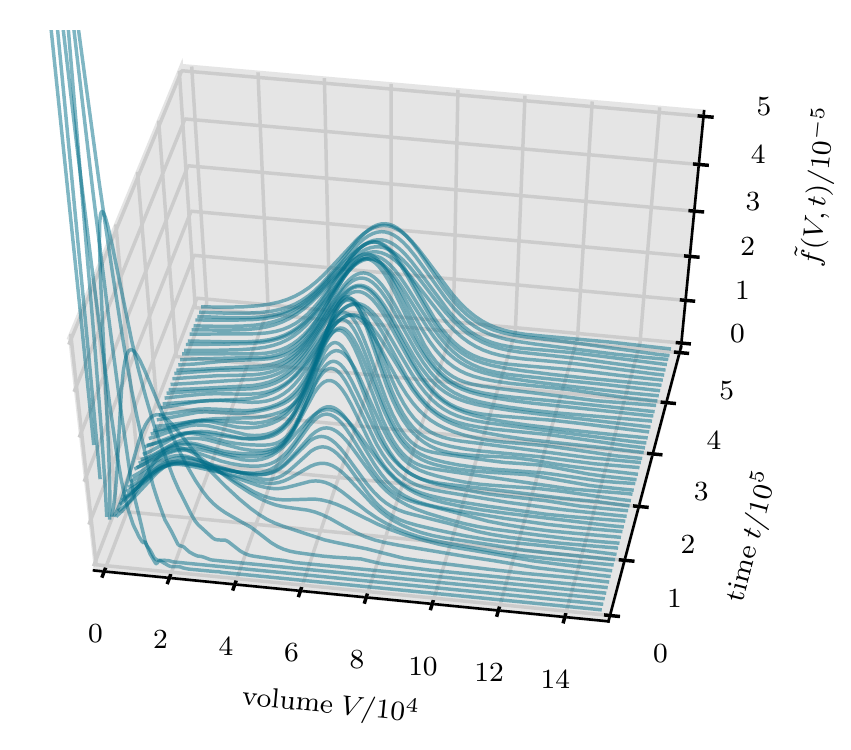}
 \caption{Time-evolution of the normalized drop size distribution $\tilde f(V,t)$ obtained at each time step by KDE for the DNS in the lower row of Fig.~\ref{fig:fig1_manysliding_timeseries}. In the early stage, the sharp peak at small drop volumes slowly broadens and shifts to the right. At later times, a second maximum at larger volumes appears, grows, broadens and becomes dominant before it finally reaches an almost steady shape.}
 \label{fig:fig2_KDE_wireframe}
\end{figure}

To quantify the coarsening process, we use the total number of drops $N_\mathrm{D}(t)$ in the domain \footnote{As the height profile $h(x,y,t)$ is a single, continuous field, we define an individual drop as a connected area $\Omega_\mathrm{drop}$ where the height exceeds a threshold slightly above the height of the adsorption layer (here $h_\mathrm{thresh} = 1.05$).}, as well as the drop size distribution $f(V,t)$ obtained by a Gaussian kernel density estimation (KDE) \footnote{To obtain $f(V,t)$, at each step of the DNS, the volume $V$ of each drop is calculated by integrating the height profile $h(x,y,t)$ over the corresponding footprint $\Omega_\mathrm{drop}$. From the resulting list at time $t$ we calculate $f(V,t)$ using a KDE \cite{scott2015multivariate}}. 
In this description, $N_{\left[V,V+\mathrm{d}V\right]}=f(V)\mathrm{d}V$ is the number of drops with a volume in the interval $\left[V,V+\mathrm{d}V\right]$. Figure~\ref{fig:fig2_KDE_wireframe} shows the time-evolution of the normalized drop size distribution $\tilde f(V,t) = f(V,t)/N_\mathrm{D}(t)$ in the inclined case of Fig.~\ref{fig:fig1_manysliding_timeseries} ($\alpha = 1.0$), while the change in the total number of drops $N_\mathrm{D}(t)$ is presented for various inclinations in Fig.~\ref{fig:fig5_statistics_vs_KDE} (bottom panel, solid lines). We find, that the inclination-induced acceleration of coarsening results in a fast drop number decrease. In particular, this coarsening is always faster than the classical rigorous scaling law $N_\mathrm{D}(t) \sim t^{-\frac{3}{4}}$ in the horizontal case \cite{otto2006coarsening}, and further accelerates with increasing inclination. In the drop size distribution (cf.~Fig.~\ref{fig:fig2_KDE_wireframe}), the fast coarsening is visible as a strong broadening and shifting of the initially tightly peaked distribution towards larger volumes. Around $t \approx 1\cdot 10^5$, it then develops a second local maximum at $V\approx6\cdot 10^4$, which grows in time at the cost of smaller values. Finally, after an inclination-dependent time $t_\mathrm{c}$, the coarsening almost stops, as indicated by a significant kink in the $N_\mathrm{D}(t)$ curve (Fig.~\ref{fig:fig5_statistics_vs_KDE} bottom) and a subsequent very slow decrease. In the DNS (see Fig.~\ref{fig:fig1_manysliding_timeseries}), this phase occurs for $t > t_c\approx 6.5\cdot 10^4$ and is characterized by drop ensembles consisting of similar-sized drops, in accordance with the quite uniform and almost stationary drop size distribution (Fig.~\ref{fig:fig2_KDE_wireframe}). Therefore, the drops slide with small relative velocities, leading to only a few coalescence events. These mergings often result in large drops that are unstable w.r.t.\ pearling and break-up again. In this way, statistically an almost stationary state is reached and kept in which merging and break up of drops balance.

\begin{figure}[htbp]
\center
 \includegraphics{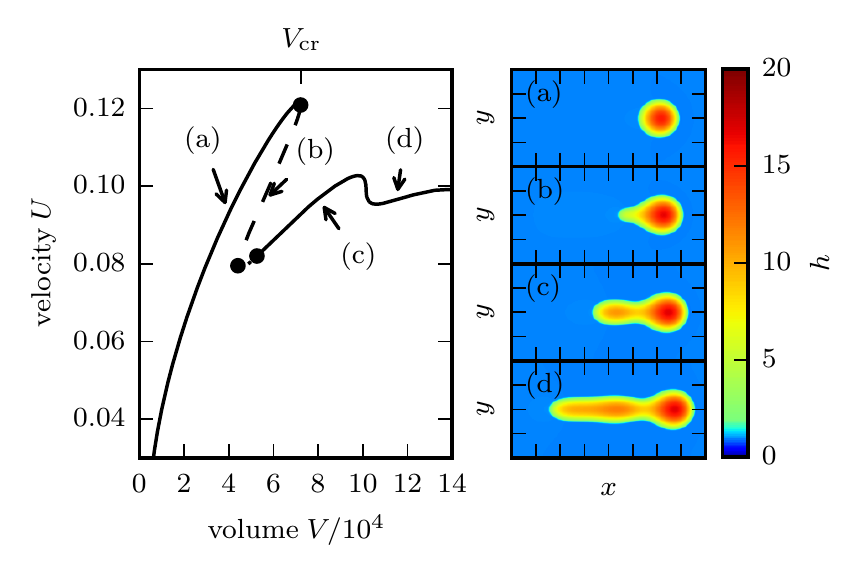}
 \caption{(left) Bifurcation diagram of sliding drops (modeled by Eq.~\eqref{eq:ThinFilmEquationSliding}) on a substrate with inclination $\alpha=1.0$. Shown is the sliding velocity $U$ in dependence on the drop volume $V$. There exist several sub-branches of stationary drops with different shape and behavior, see labels (a) to (d) and the corresponding solutions on the right.}
 \label{fig:fig3_bifurcation}
\end{figure}

\paragraph{Stability Properties of Single Drops}
The time evolution of the drop size distribution results from the interplay of drop interactions (dominated by their relative velocity) and stability properties of individual drops. Both information is presented in Fig.~\ref{fig:fig3_bifurcation} in the form of a bifurcation diagram obtained by pseudo-arclength continuation within the PDE2Path framework \cite{Uecker2014}. It shows for a single drop at fixed inclination the dependence of sliding velocity on drop volume $V$ (cf.\ Ref.~\cite{EWGT2016PRF} for other cases and implementation details).
Figure \ref{fig:fig3_bifurcation} reveals the existence of a variety of different drop shapes, velocities and stability properties. For small drop volumes $V$, only simple, almost ellipsoidal cap-shaped drops exist (sub-branch (a)). Increasing $V$, this sub-branch terminates at a critical volume $V_\mathrm{cr}\approx 7.2 \cdot 10^4$ in a saddle-node bifurcation, which also connects it to sub-branch (b), whose drops exhibit an elongated tail and are linearly unstable. Sub-branch (b) connects to the stable sub-branches (c,d) via another saddle-node bifurcation and a subsequent Hopf bifurcation (cf.~\cite{EWGT2016PRF}). Although at small volumes only the drops of sub-branch (a) exist, from $V \approx 5.3\cdot 10^6$ onwards, we find a multistability of sub-branch (a) and the elongated drops of sub-branch (c). In the DNS, one mainly observes drops from sub-branch (a) because drops of larger volume than $V_\mathrm{cr}$ which are formed by merging are normally unstable and decay by pearling. However, sometimes the merged drop is elongated and linearly stable, i.e., on sub-branches (c,d).

\begin{figure}[htbp]
\center
 \includegraphics{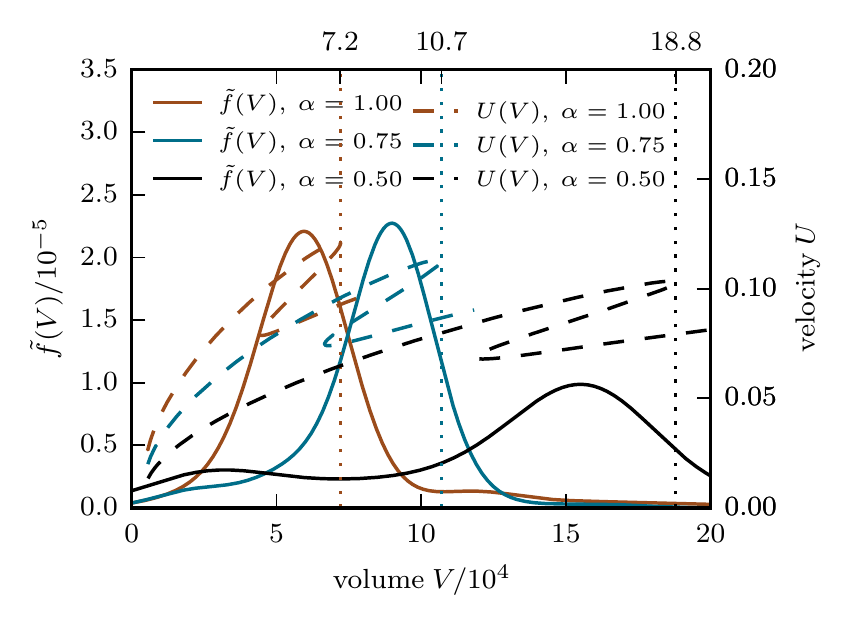}
 \caption{Comparison of late-stage quasi-stationary normalized drop size distributions $\tilde f(V)$ at $t=5\cdot10^5$ and bifurcation diagrams $U(V)$ for single drops at different fixed inclinations $\alpha$. The vertical dotted lines indicate the position $V_\mathrm{cr}$ of the respective saddle-node bifurcation.}
 \label{fig:fig4_KDE_vs_bif}
\end{figure}

\begin{figure}[ht]
\center
\includegraphics{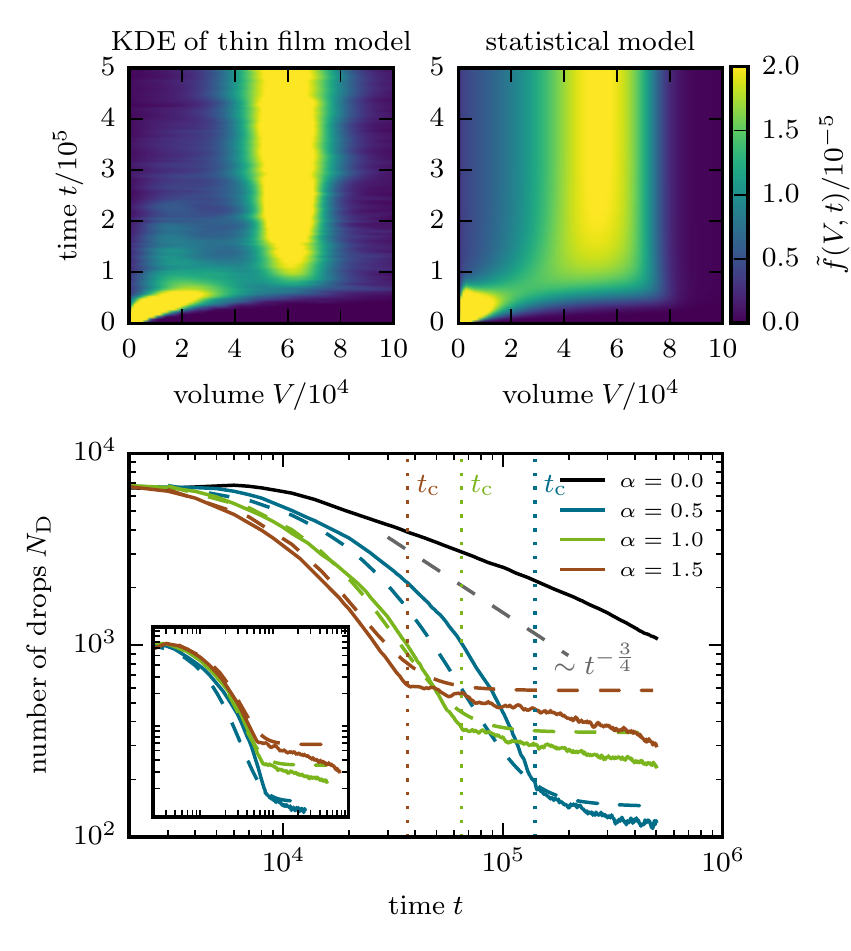}
 \caption{(top) Space-time plots show the time evolution of the drop size distribution $\tilde f(V,t)$ at $\alpha=1.0$ obtained by KDE from (left) a DNS of Eq.~\eqref{eq:ThinFilmEquationSliding} (cf.~Fig.~\ref{fig:fig2_KDE_wireframe}) and (right) a simulation of the statistical model \eqref{eq:StatisticalModel}. (bottom) Comparison of the time evolution of the drop number $N_D(t)$ for different inclinations $\alpha$ as obtained by DNS (solid lines) and statistical model (dashed lines). The coarsening with inclination is always faster than the rigorous upper bound $N_D(t) \sim t^{-3/4}$ \cite{otto2006coarsening}. The inset shows the same data with the $t$-axis scaled by $\alpha$, which results in a master curve in the collision-dominated regime.
The remaining parameters are $j = 6.25\cdot 10^{-6}$, $k_1=\frac{1}{30}$, $b_\mathrm{V} = V_\mathrm{cr}/10$.} \label{fig:fig5_statistics_vs_KDE}
\end{figure}

Next, we connect the information gained from the bifurcation study of the individual drop to the ensemble dynamics. As at relatively low inclinations stable elongated drops are rarely formed, we focus on sub-branch (a): The bifurcation point at $V_\mathrm{cr}$ provides the stability limit for simple drops and, therefore, sets an upper limiting volume for the ensemble DNS. Figure \ref{fig:fig4_KDE_vs_bif} shows bifurcation curves together with the late-stage quasi-stationary drop size distributions obtained from DNS. Comparison shows, that the location of the main peak of the distribution is directly connected to the position of the saddle-node bifurcation at $V_\mathrm{cr}$: The number of drops with $V>V_\mathrm{cr}$ decreases significantly. Indeed, $V_\mathrm{cr}$ almost coincides with the r.h.s.\ inflection point of the size distribution. These observations equally hold for different inclinations $\alpha$ (see Fig.~\ref{fig:fig4_KDE_vs_bif}).

Further, we extract from bifurcation diagrams as the one in Fig.~\ref{fig:fig3_bifurcation} power laws that relate (i) drop velocity and volume at fixed inclination 
\begin{equation}
U(V) = a_0 \alpha V^{\beta_0},
\label{eq:SpeedScaling}
\end{equation}
and (ii) the critical $V_\mathrm{cr}$ and inclination,
\begin{equation}
V_\mathrm{cr} = a_1 \alpha^{\beta_1},
\label{eq:VolumeScaling}
\end{equation}
with $a_0 = 2.2\cdot 10^{-4}$, $\beta_0 = 0.569$, $a_1 = 7.18\cdot10^4$ and $\beta_1 = -1.40$ \cite{EWGT2016PRF}.

\paragraph{Statistical Model}
In the final step, the obtained 'single-drop' information is employed to develop a minimal statistical model for the ensemble dynamics as characterized by the unnormalized drop size distribution $f(V,t)$. To capture the coarsening dynamics that is dominated by the interplay of collision-caused merging and instability-caused splitting of drops, we extend Smoluchowski's continuous rate equation for coagulation \cite{Smoluchowski1916}, following the approach of Meakin et al. for breath figures (see \cite{meakin1992droplet} and references therein). Thereby, loss and gain of drops of each volume are accounted for through continuous transition rate kernels for coalescence and fragmentation. The model conserves the total volume and reads
\begin{align}\label{eq:StatisticalModel}
 &\partial_t  f(V,t)= -\underbrace{\int\limits_0^\infty K(V,\tilde V)\, f(V)\, f(\tilde V) \, \mathrm{d}\tilde V}_{\mathrm{loss~due~to~coalescence}} \nonumber \\
&+\underbrace{\int\limits_0^V \frac{1}{2}K(\tilde V,V-\tilde V) \,  f(\tilde V) \,  f(V-\tilde V)\,\mathrm{d}\tilde V \, }_{\mathrm{gain~due~to~coalescence}}\nonumber\\
&-\underbrace{\int\limits_0^\infty \frac{1}{2}J(V,\tilde V)\, f(V)\,\mathrm{d}\tilde V}_{\mathrm{loss~due~to~fragmentation}}+\underbrace{\int\limits_0^\infty J(\tilde V,V)\, f(\tilde V)\,\mathrm{d}\tilde V}_{\mathrm{gain~due~to~fragmentation}},
\end{align}
\begin{align}
\label{eq:KernelK}
 \mathrm{where~~~}K(V_a,V_b) = \frac{2k_1}{L} \left| U(V_a) - U(V_b) \right|,
\end{align}
\begin{align}
 J(V_a,V_b) = j\ \sigma(V_a,V_\mathrm{cr})\ \Theta \left(V_a - V_b\right).
 \label{eq:KernelJ}
\end{align}
The properties of the kernels $K$ and $J$ in this non-local
evolution equation are crucial features of the coarse-grained model. We deduce them from the single-drop results \eqref{eq:SpeedScaling} and \eqref{eq:VolumeScaling} above and employ a minimum of free parameters and assumptions. In particular, the kernel $K(V_a,V_b)$ (cf.~Eq.~\eqref{eq:KernelK}) accounts for the coalescence of two drops with volumes $V_a$ and $V_b$. It sets the frequency of collisions as the ratio of the relative drop velocity and the mean distance $L/2$ between two drops on the domain.
The drop velocities $U(V)$ are given by the obtained scaling law \eqref{eq:SpeedScaling} with the only a priori unknown parameter being $k_1$. It is a measure for the reduction of the number of collisions because all drops slide in the same direction and therefore only interact with a subset of the other drops. The other kernel $J(V_a,V_b)$ (cf.~Eq.~\eqref{eq:KernelJ}) with the sigmoid function $\sigma(V_a,V_\mathrm{cr})$ \footnote{We use $\sigma(V_a,V_\mathrm{cr}) = \frac{1}{2} \left( 1+ \tanh \left( \frac{V_a-(V_\mathrm{cr}+2b_V)}{b_V}\right)\right)$} accounts for drop splitting and corresponds to the simplest implementation of the instability threshold obtained above. In particular, drops with $V_a>V_\mathrm{cr}$ fragment into two drops of volume $V_b$ and $V_a-V_b$, respectively, with equal probability for all $V_b<V_\mathrm{cr}$ (expressed by the Heaviside function $\Theta$) \footnote{This is a simple rational choice as information about the specific instability
timescales for specific drop sizes and particular fragmentation ratios is difficult to obtain and too complex for a simple model \cite{EWGT2016PRF}.}. Here, the free parameters are the smoothness $b_V$ of the transition to the unstable regime and the timescale ratio $j$ between fragmentation and coalescence processes. 

The developed statistical model \eqref{eq:StatisticalModel} is solved numerically \footnote{Discretized on the volume domain and with a 4th order Runge-Kutta time-stepping scheme.} employing initial conditions corresponding to early stages of the DNS of Eq.~\eqref{eq:ThinFilmEquationSliding} (e.g., in Fig.~\ref{fig:fig1_manysliding_timeseries}). As a result, the top row of Fig.~\ref{fig:fig5_statistics_vs_KDE} compares the two dynamics of the drop size distribution as measured in the DNS of the thin film equation \eqref{eq:ThinFilmEquationSliding} and in the simulation of the statistical model \eqref{eq:StatisticalModel}. It shows a very good agreement of all main features, as e.g., the appearence of a second peak and the convergence to a quasi-stationary distribution.

Furthermore, the bottom panel of Fig.~\ref{fig:fig5_statistics_vs_KDE} compares the evolution of the drop number $N_D$ for different inclination angles $\alpha$
fixing in all cases parameters $k_1$ and $j$ such that a best fit results for simulations with $\alpha=1.0$. Nevertheless, the predictions of the statistical model for all inclinations agree very well with the DNS results. This gives clear evidence that the dynamics of the ensemble properties resulting from many individual complex coalescence and fragmentation processes can be rather well captured by our simple statistical model.

\paragraph{Conclusions}
We have investigated the coarsening behavior of ensembles of interacting sliding drops employing a thin-film equation. We have shown in direct simulations that a balance of coalescence and fragmentation processes emerges that can be related to stability properties of individual drops as captured in a single-drop bifurcation diagram. Dynamically, the statistical ensemble properties converge to an almost stationary state. Further, based on the gained single-drop information, we have developed a minimal statistical model that faithfully captures the main ensemble dynamics and very well compares to the full direct numerical simulations. We believe that the proposed methodology of employing 'microscopic' information in the form of bifurcation properties of individual entities (here drops), to derive coarse-grained 'macroscopic' statistical models for the ensemble dynamics, represents a multi-scale approach that will prove useful in other nonlinear nonequilibrium systems.\par
We acknowledge partial support by DFG within the Sino-German Collaborative Research Center TRR 61, and GIF under grant I-1361-401.10/2016.
\bibliography{literature_Wilczek2017PRL}
\end{document}